\documentclass[twocolumn,prb,superscriptaddress,floatfix]{revtex4}
\usepackage{color}
\usepackage{colordvi}
\usepackage{amsmath}
\usepackage{amssymb}
\usepackage{fancybox}
\usepackage{epsfig}
\usepackage{graphicx}
\begin{document}

\title{Magnetic properties of interacting, disordered electron systems in $d=2$ dimensions}

\author{Prabuddha B. Chakraborty}
\email{prabuddha.chakraborty@physik.uni-augsburg.de}
\affiliation{ Theoretical Physics III, Center for Electronic
Correlations and Magnetism, Institute of Physics, University of
Augsburg, D-86135, Augsburg, Germany}
\author{Krzysztof Byczuk}
\affiliation{Institute of Theoretical Physics, Faculty of Physics,
University of Warsaw, ul. Ho\.za 69, 00-681, Warszawa, Poland}
\author{Dieter Vollhardt}
\affiliation{ Theoretical Physics III, Center for Electronic
Correlations and Magnetism, Institute of Physics, University of
Augsburg, D-86135, Augsburg, Germany}

\begin{abstract}
We compute the magnetic susceptibilities of interacting electrons in the presence of disorder on a two-dimensional square lattice by means of quantum Monte Carlo simulations. Clear evidence is found that at sufficiently low temperatures disorder can lead to an enhancement of  the ferromagnetic susceptibility. We show that it is not related to the transition from a metal to an Anderson insulator in two dimensions, but is a rather general low temperature property of interacting, disordered electronic systems.     
\end{abstract}
\date{\today}

\maketitle

\section{Introduction}

The interplay of disorder and Coulomb repulsion between electrons raises many fundamental questions not only in solid state physics\cite{Lee85, Altshuler85, Belitz94, Anderson50-2010} but also in the physics of cold atoms.\cite{Bloch:2008, Aspect:2009, Sanchez-Palencia:2010} Electron repulsion can lead to a Mott insulator,\cite{Mott90} a description of which must invoke many-body physics. On the other hand, presence of disorder in a system of non-interacting electrons can cause Anderson localization.\cite{Anderson58, Anderson50-2010} The simultaneous presence of disorder and electronic interaction is known to give rise to new, subtle many-body phenomena.\cite{Lee85, Altshuler85, Belitz94,Anderson50-2010} 
An important question in the physics of interacting and disordered electrons is the stability of magnetic order. The effect of disorder has on magnetic long-range order has been explored by various analytical \cite{Spin_diffusion, Finkelshtein83} and numerical methods.\cite{Ulmke:1997, Ulmke:1995, Byczuk:2003, Byczuk:2009, Tusch:1993, Heidarian:2004} A microscopic lattice model to study these effects is the Anderson-Hubbard model, which has been explored using quantum Monte Carlo (QMC),\cite{Ulmke:1997} dynamical mean-field theory (DMFT) \cite{Ulmke:1995, Byczuk:2003, Byczuk:2009} and Hartree-Fock calculations.\cite{Tusch:1993, Heidarian:2004} One of the questions that have received particular attention is whether the long-range antiferromagnetic order (AFLRO), that occurs at half-filling in the clean limit, is stable when disorder is present. Increasing disorder has generically been found to be detrimental for AFLRO.\cite{Ulmke:1997, Byczuk:2009} 

Magnetic instabilities and the formation of local moments have also been investigated in high mobility semiconductor heterostructures.\cite{Kravchenko:1994} Some experiments have found the ferromagnetic (uniform) susceptibility to behave critically near the MIT,\cite{Shashkin:2001} while other experiments only found a strong enhancement.\cite{Pudalov:2002, Prus:2003}

Analytical work has predicted critical behavior of the ferromagnetic susceptibility as the metal-insulator transition is approached from either side of the transition, and the physics has been attributed to either the formation of local moments \cite{Paalanen:1986, Castellani:1984} or to the vanishing of the spin diffusion constant either at or near the (charge)transport critical point.\cite{Spin_diffusion, Punnoose:2005}

QMC investigations in two dimensions have also suggested the existence of a quantum critical point (QCP) between a metallic and an (Anderson)insulating phase when the disorder strength is varied.\cite{Denteneer:1999, Denteneer:2001, Denteneer:2003, Chakraborty:2007, Chakraborty:2010} Magnetic properties both near and away from the QCP have received less attention. Denteneer {\it{et al.}} \cite{Denteneer:1999} found an increase in the ferromagnetic susceptibility both sides of the metal-insulator transition and this behaviour was attributed to the formation of local moments.\cite{Paalanen:1986, Castellani:1984}

In this paper, we report results on the magnetic properties of the Anderson-Hubbard model in two dimensions obtained through extensive QMC simulations. We concentrate on the ferromagnetic (FM) and the antiferromagnetic (AF) susceptibilities. There are two main results: (i) We find that, at sufficiently low temperatures, an increase in the disorder increases the FM susceptibility, and (ii) the enhancement of the FM susceptibility is related neither to the MIT, nor can it be explained by the formation of local moments. In contrast, increasing disorder always reduces the AF susceptibility.

\section{Correlation functions for the Anderson-Hubbard Model in $d=2$}
Our investigation of interacting electrons in the presence of disorder is based on the Anderson-Hubbard Hamiltonian on a square lattice
\begin{equation}
H = T\{\epsilon_i\}+U\sum_i n_{i\uparrow}n_{i\downarrow}.
\label{Hamiltonian}
\end{equation}
Here
\begin{eqnarray}
T\{\epsilon_i\}&=&-t\sum_{<ij>\sigma}
c_{i\sigma}^\dagger c_{j\sigma}
+\sum_{i \sigma}(\epsilon_i - \mu)n_{i\sigma},
\label{quadratic}
\end{eqnarray}
is the single-electron part where $c_{i\sigma}^\dagger$ ($c_{i\sigma}$) are
fermion creation (annihilation) operators for site $\textbf{R}_i$ and spin $\sigma$,
$n_{i\sigma}=c_{i\sigma}^\dagger c_{i\sigma}$ is the operator for the local density, $\mu$ denotes the chemical potential, and $t$ is the hopping amplitude for electrons between nearest neighbor sites.
%
The local energies $\epsilon_i$ are random variables which are sampled uniformly from the
interval $[-\Delta/2,\Delta/2]$; hence the width $\Delta$ characterizes the strength of the disorder. The interaction is taken to be repulsive ($U>0$).

The model is solved numerically using determinantal QMC (DQMC).\cite{Blankenbecler:1981} The hopping integral $t$ sets the unit of energy (we set $\hbar$=$k_{B}$=1) and the simulation now contains three independent energy-scales: the disorder strength $\Delta$, the interaction strength $U$, and the temperature $T$. The average filling $n$ is tuned by the chemical potential $\mu$. 

To extract the magnetic behaviour, we focus on on the real-space equal-time spin-spin correlation function,\cite{Varney:2009} defined as
\begin{equation}
C(\textbf{r}) = \langle S^{z}(\textbf{R}_{i}+\textbf{r})S^{z}(\textbf{R}_{i})\rangle,
\label{definition_realspace}
\end{equation}    
where $S^z(\textbf{R}_{i})=n_{i\uparrow}-n_{i\downarrow}$. $C(\textbf{r})$ measures the extent to which the z-component of a spin on site $\textbf{R}_{i}$ is correlated to the z-component of another spin at a distance $\textbf{r}$. Although we have used the operator $S^{z}$ in Eq.~\ref{definition_realspace} to define $C(\textbf{r})$, we explicitly confirmed that our calculations have the full SU(2) invariance of the problem. 

From $C(\textbf{r})$, we define the spin structure factor $S(\textbf{q})$ as the Fourier transform:
\begin{equation}
S(\textbf{q}) = \sum_{\textbf{r}}e^{i\textbf{q}.\textbf{r}}C(\textbf{r}).
\end{equation}
The $\textbf{q}$-dependent susceptibility is obtained from $\chi(\textbf{q})=\beta S(\textbf{q})$.\cite{Denteneer:1999} The wave-vectors $\textbf{q}=(0,0)$ yield the FM susceptibility $\chi_{F}$, and $\textbf{q}=(\pi,\pi)$ yield the AF susceptibility $\chi_{AF}$, respectively. The lattice spacing $a$ is assumed to be unity. All plots shown are for a $10 \times 10$ square lattice. We have confirmed that the conclusions are unchanged for computations on $8 \times 8$ and $12 \times 12$ square lattices by explicit simulations.

There are two sources of statistical error in this analysis: one is due to  the QMC simulations, the other arises from disorder averaging. For all parameter sets studied here, the intrinsic QMC error for any given disorder realization is much smaller than the error arising from different disorder realizations. 

\section{Results}

\begin{figure}
\centering
\begin{tabular}{c}
\epsfig{file=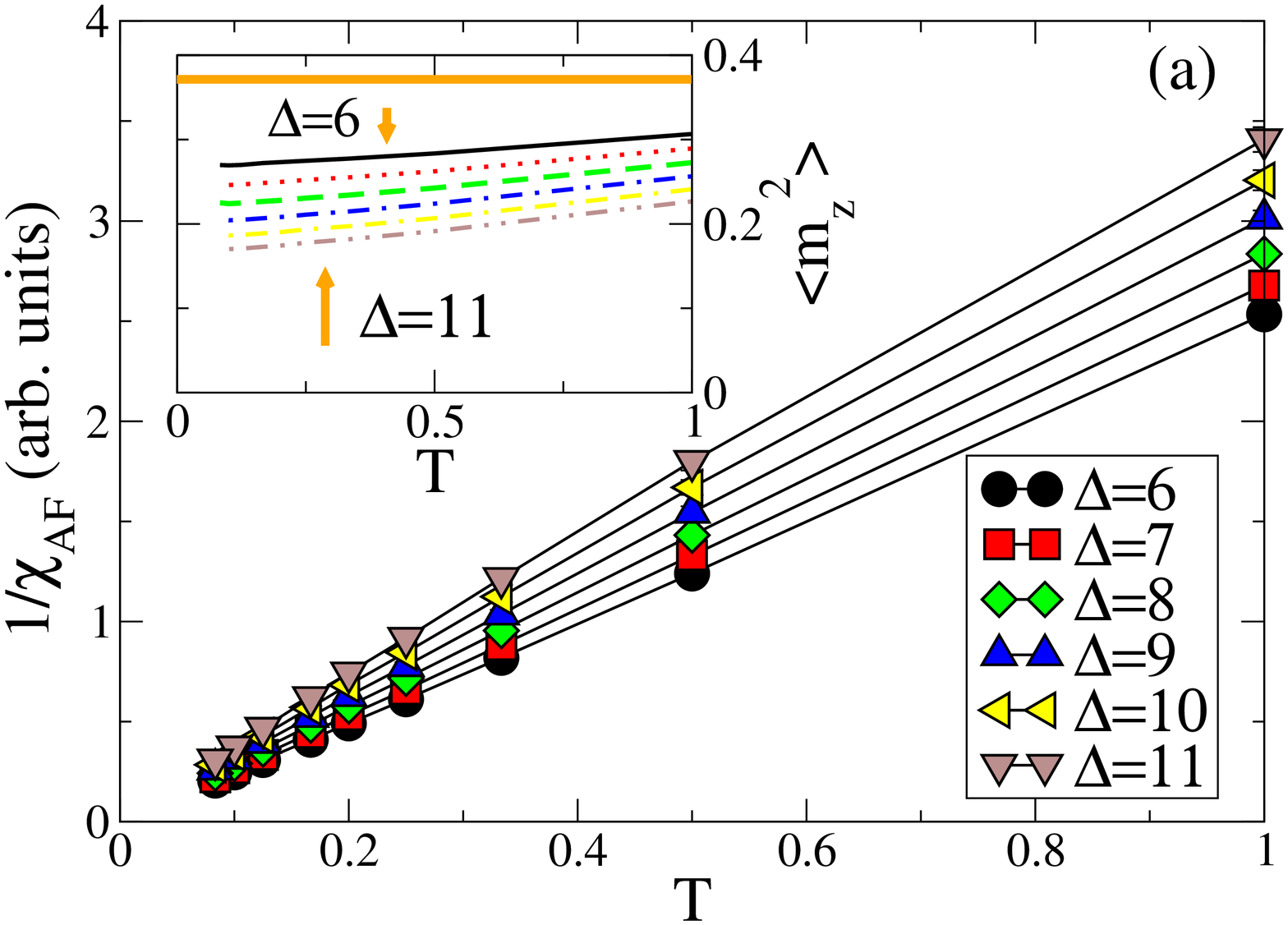,width=0.7\linewidth,clip=true,angle=-90} \\
\epsfig{file=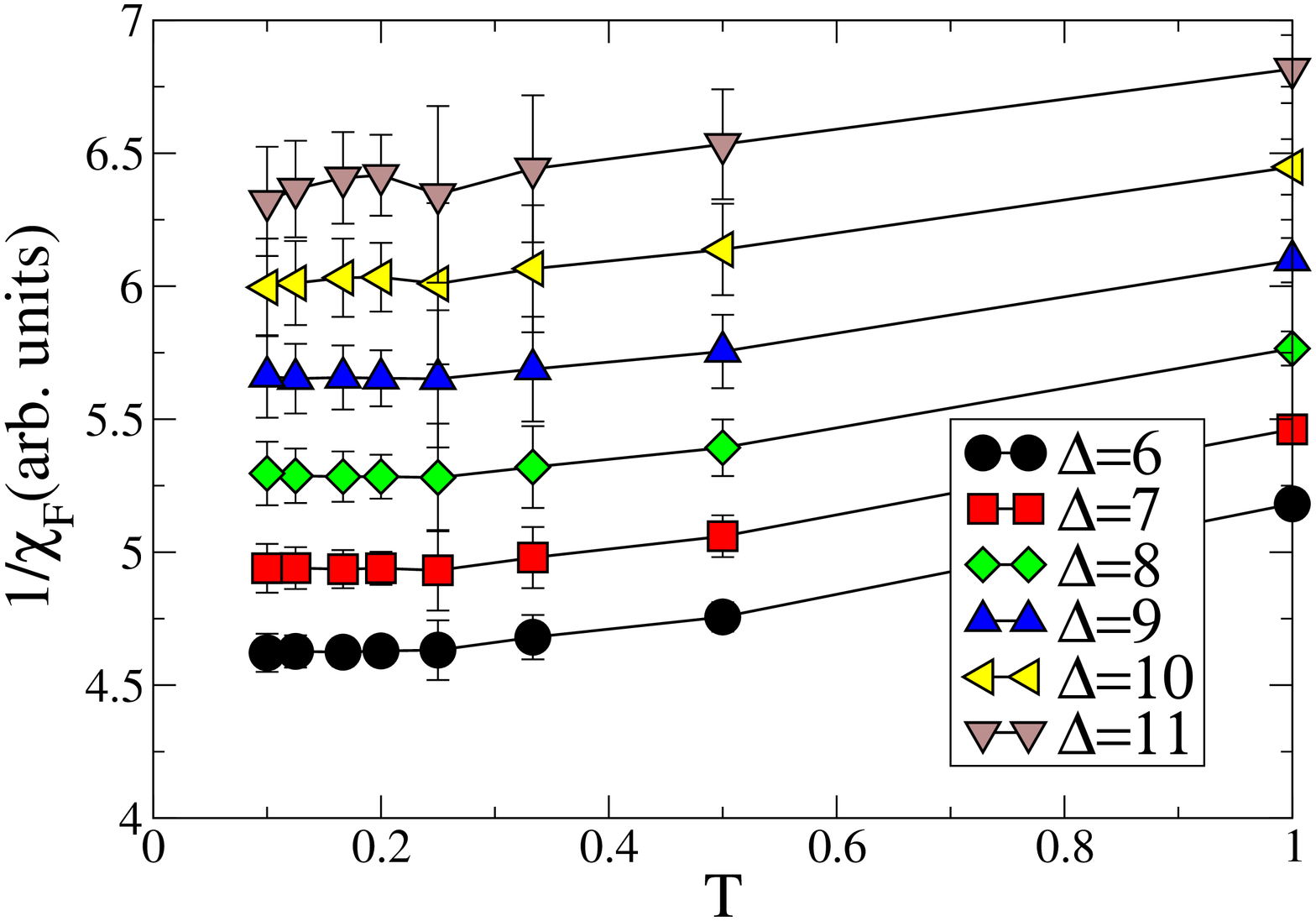,width=0.7\linewidth,clip=true,angle=-90}
\end{tabular}
\caption{(Color online) (a) Inverse AF susceptibility and, (b) inverse ferromagnetic susceptibility computed for a 10 $\times$ 10 square lattice at quarter-filling ($n=0.5$) for $U=0$. The susceptibility data is averaged over 10 disorder realizations for the highest temperatures ($T=1.0, 0.5, 0.333, 0.25)$, 80 disorder realizations for intermediate temperatures ($T=0.2, 0.167$), 100 disorder realizations for low temperatures ($T=0.125, 0.1$) and 120 disorder realizations for the lowest temperature ($T=0.0833$). {\emph{Inset}}: Square of the local magnetic moment {\emph{vs}} T from the same datasets. The horizontal line is the local moment squared ($=0.375$) when $\Delta=0$.}
\label{U0_AF_uniform}
\end{figure}

We start with the non-interacting system. In Figs.~\ref{U0_AF_uniform}(a) and (b) we show the dependence of the inverse AF and inverse FM susceptibilities, respectively, on $T$ for $U=0$ and various disorder strengths. Without interaction, an increase in disorder always reduces both the AF and FM susceptibilities. Namely, increasing disorder creates more fluctuations in the site-energy landscape and, since $U=0$, two electrons of opposite spins will tend to occupy the lowest energy sites, thereby reducing the effective local moment and destroying magnetism. The inset plots the square of the local moment as a function of temperature, defined as $\langle m^2_{z,i}\rangle = C(\textbf{r}=0) = \langle n_{i}\rangle - 2\langle n_{i\uparrow}n_{i\downarrow}\rangle $.     

\begin{figure}
\centering
\begin{tabular}{c}
\epsfig{file=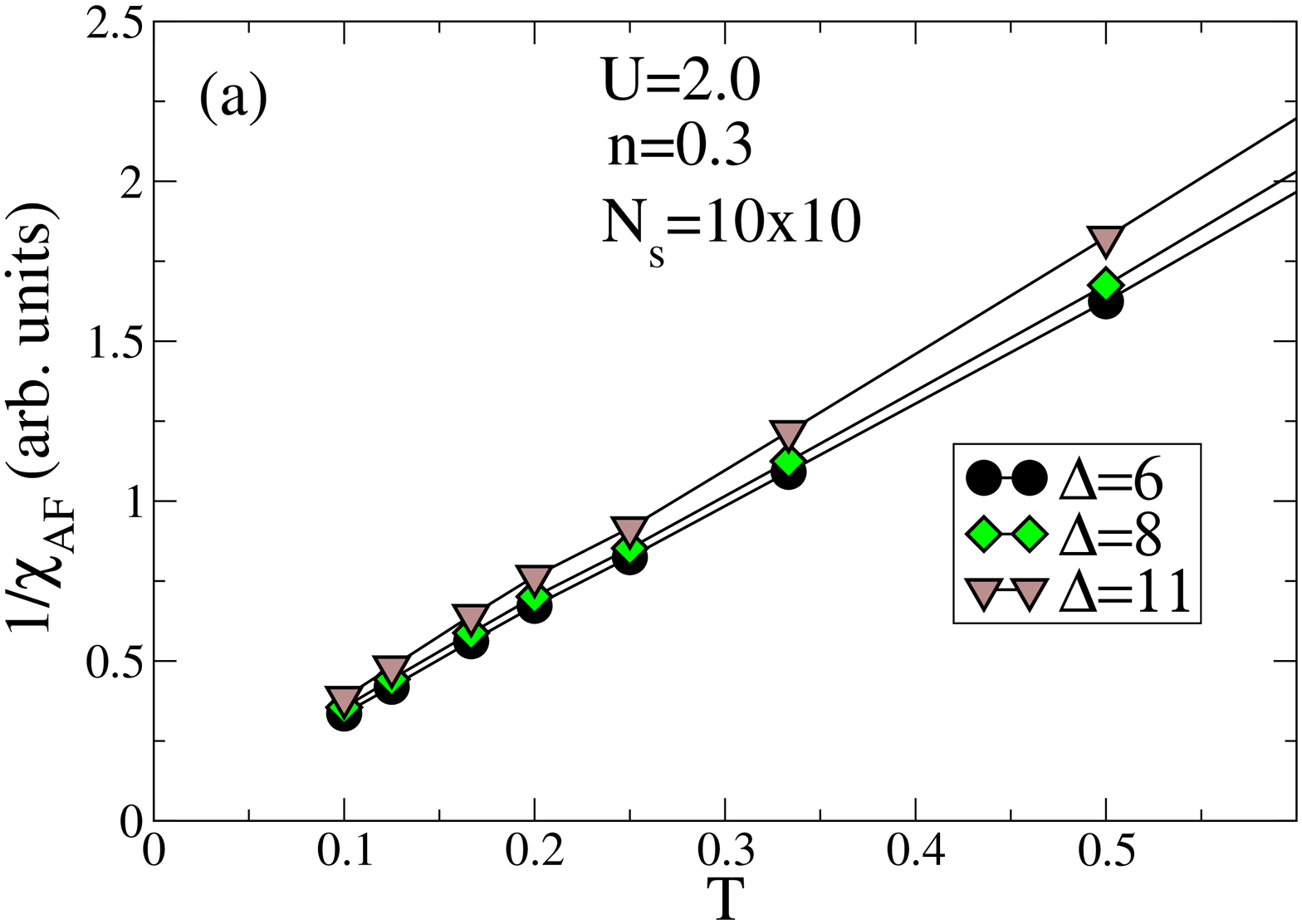,width=0.7\linewidth,clip=true,angle=-90} \\
\epsfig{file=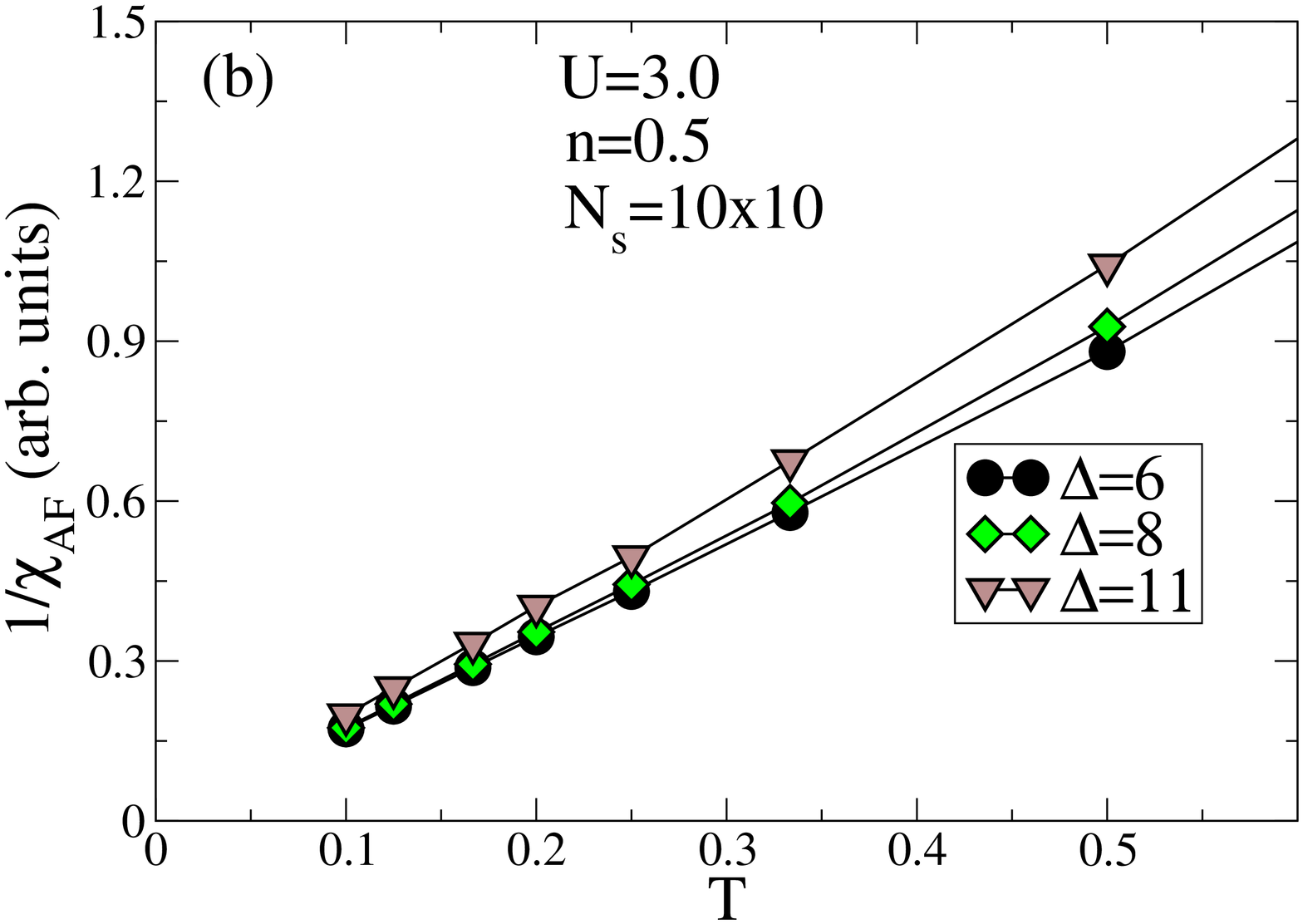,width=0.7\linewidth,clip=true,angle=-90}
\end{tabular}
\caption{(Color online) Inverse of the AF susceptibility for the parameter sets (a) $U=2$, $n=0.3$ and (b) $U=3$, $n=0.5$. The number of disorder realizations is the same as in Fig. \ref{U0_AF_uniform}. Similar behaviour is observed for all other parameter sets investigated.   }
\label{inv_afchi_two_param_sets}
\end{figure}

Now we include the interaction $U$. In Fig.~\ref{inv_afchi_two_param_sets}, we plot the inverse AF susceptibility for two different $(U,n)$ parameter sets. In all cases, $1/\chi_{AF}$ decreases linearly with decreasing $T$. It is clearly seen that the plot for every value of $\Delta$ can be extrapolated to zero at $T=0$, which signifies that there will be no AF ordering. This behaviour is expected for the Anderson-Hubbard model far away from half-filling ($n=1$). Furthermore, the AF susceptibility decreases monotonically as the disorder strength is increased.

\begin{figure}
\centering
\begin{tabular}{c}
\epsfig{file=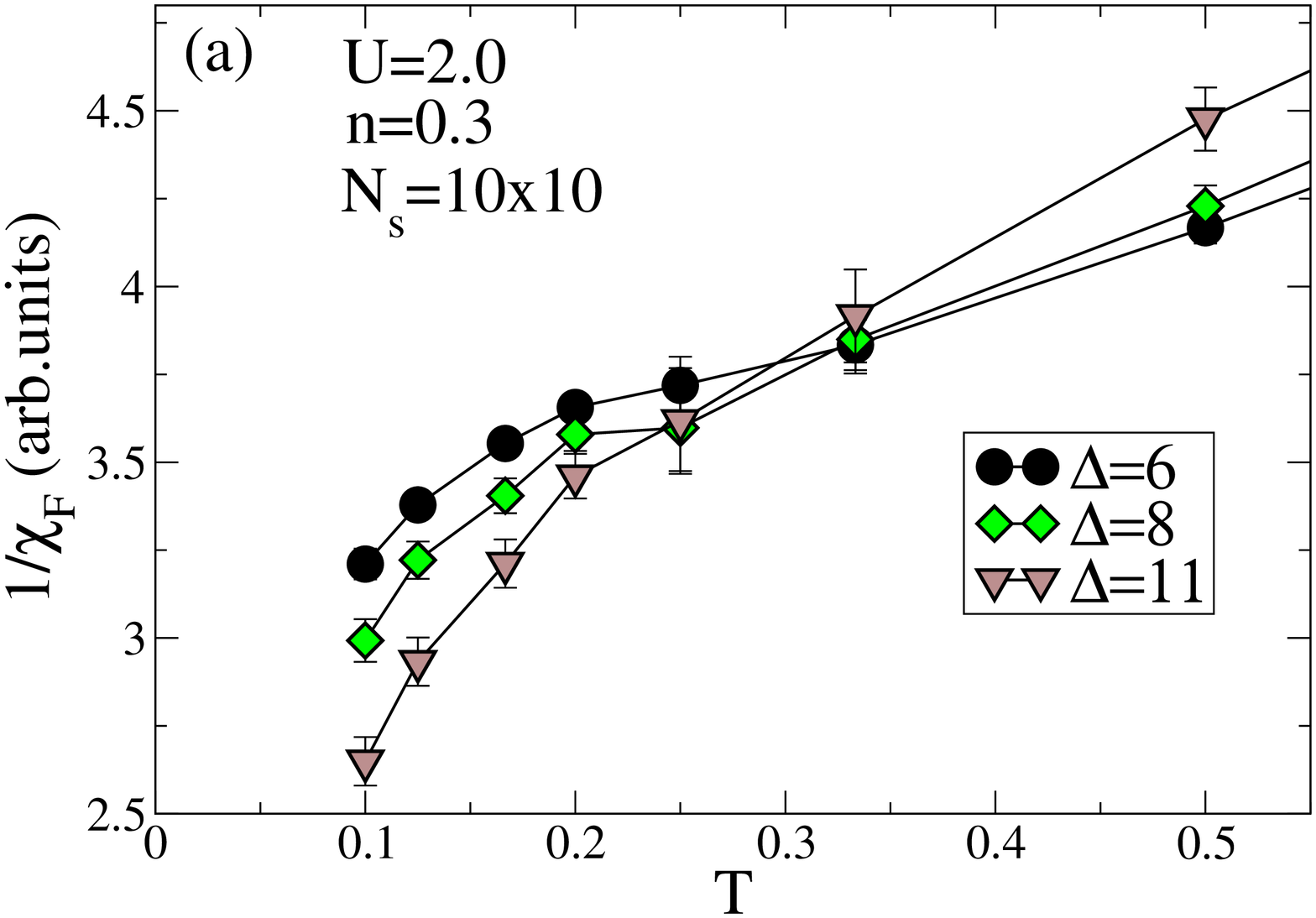,width=0.7\linewidth,clip=true,angle=-90} \\
\epsfig{file=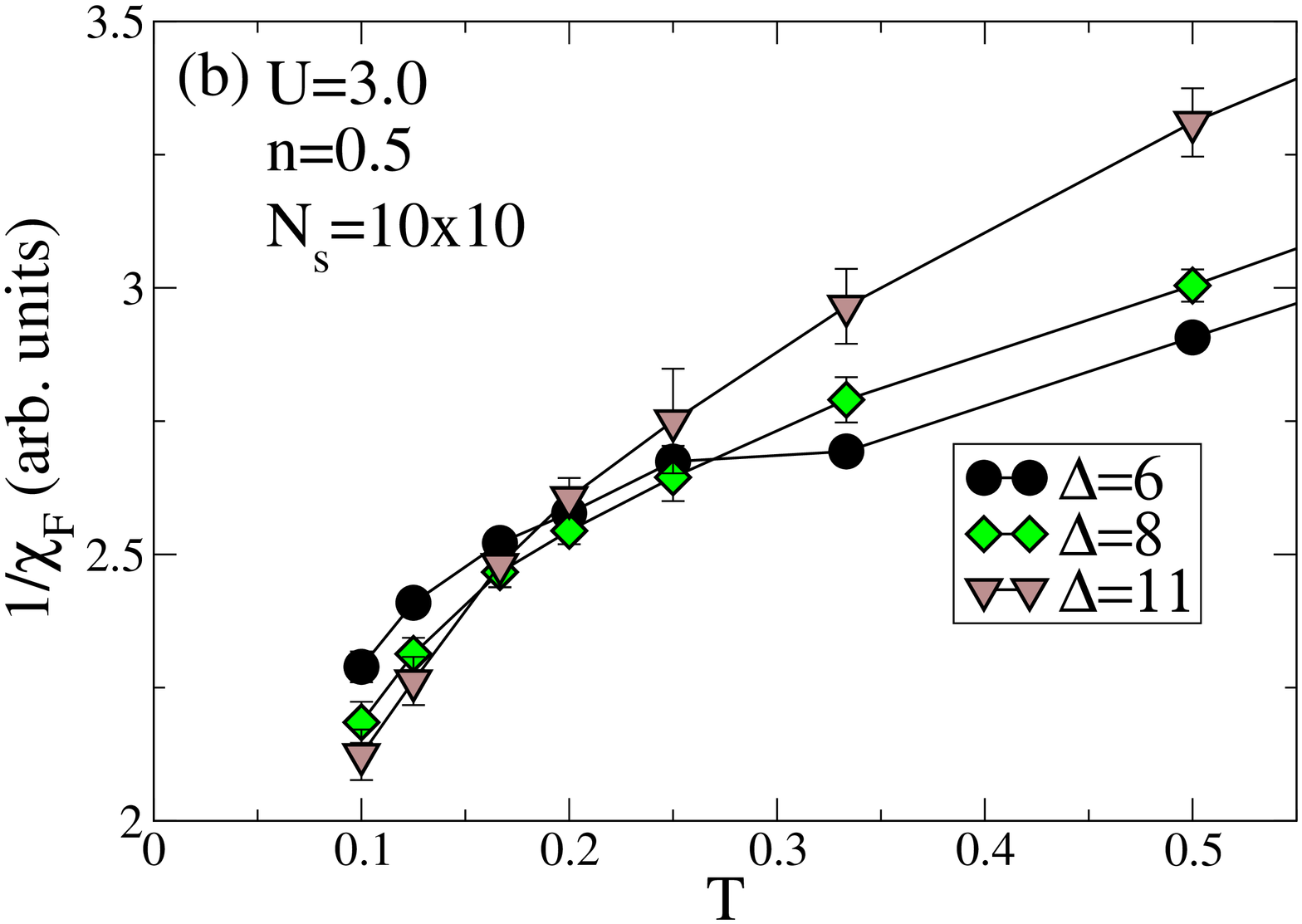,width=0.7\linewidth,clip=true,angle=-90}
\end{tabular}
\caption{(Color online) Inverse of the ferromagnetic susceptibility for the parameter sets (a) $U=2$, $n=0.3$ and (b) $U=3$, $n=0.5$. At sufficiently low $T$ ($T$ less than about $0.3$ in (a)) $\chi_{F}$ increases with increasing disorder strength. }
\label{inv_ferrochi_two_param_sets}
\end{figure}

This is not surprising since AFLRO will be hindered by randomness. Namely, electrons inevitably find it difficult to hop (and thus gain kinetic energy) when the energies on two neighbouring sites are very different. Ferromagnetism, on the other hand, stems from the need to avoid strong Coulomb repulsion locally. In Fig.~\ref{inv_ferrochi_two_param_sets}, we plot $1/\chi_{F}$ {\emph{vs}}. $T$ for the same two parameter sets as in Fig.~\ref{inv_afchi_two_param_sets}. At high temperatures $\chi_{F}$ decreases with increasing disorder strength. However, at low temperatures, the opposite behaviour is seen to occur. We point out the difference between Fig.~\ref{inv_ferrochi_two_param_sets} and the non-interacting case in Fig.\ref{U0_AF_uniform}(b). Thus it is clear that the enhancement of $\chi_{F}$ with increasing disorder at low temperatures takes place only in the simultaneous presence of interaction and disorder. Also, this behaviour is ubiquitous in all parameter sets we have investigated. Even at weak coupling (e.g. $U=1$), the ferromagnetic susceptibility rises with increasing disorder at lower temperatures.   

\begin{figure}
\includegraphics[clip=true,angle=-90,width=0.5\textwidth]{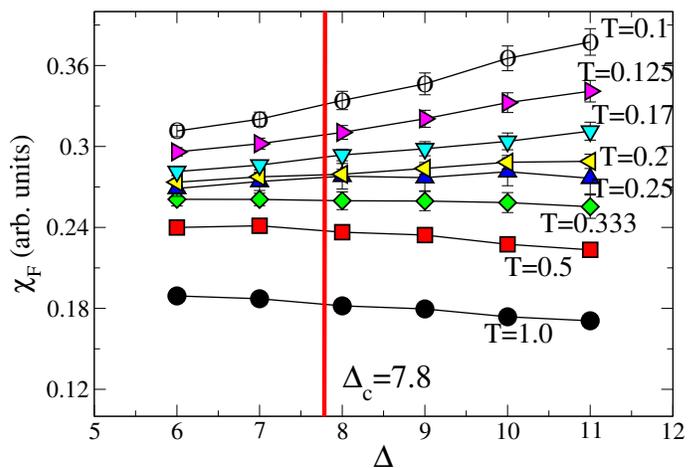}
\caption{(Color online) Ferromagnetic susceptibility as a function of disorder strength from high (bottom)
to low (top) temperatures for $U=2$, $n=0.3$.  }
\label{ferrochi_vs_D}
\end{figure}

To investigate if the enhancement of the ferromagnetic susceptibility is related to the metal-insulator transition (MIT) in $d=2$, we plot $\chi_{F}$ in Fig.~\ref{ferrochi_vs_D} against the disorder strength for $U=2$, $n=0.3$. For these parameters, the $d=2$ Anderson-Hubbard model shows a metal-(Anderson) insulator quantum phase transition at $\Delta_c=7.8$.\cite{Chakraborty:2010} It is clear from Fig.~\ref{ferrochi_vs_D} that $\chi_{F}$ does not show any enhancement close to the MIT. This holds for all parameter sets we have investigated.

\begin{figure}
\includegraphics[clip=true,angle=-90,width=0.5\textwidth]{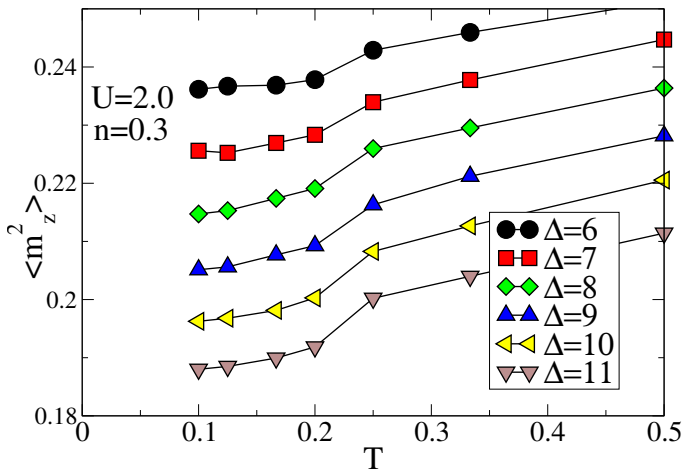}
\caption{(Color online) Square of the average moment {\emph{vs}}. temperature for different disorder strengths for $U=2$, $n=0.3$. The number of disorder realizations is the same as Fig.~\ref{U0_AF_uniform}. The magnetic moment continues to monotonically decrease as disorder is increased for all $T$. The temperature values where we observe a sudden change in the slope of $\langle m_{z}^{2}\rangle$ for all disorder strengths correspond closely to the temperature values which demarcate high and low temperature regimes in Fig.~\ref{inv_ferrochi_two_param_sets}(a). }   
\label{musq}
\end{figure}

In Fig.~\ref{musq} we plot the average squared moment {\emph{vs}}. temperature for different values of the disorder strength for $U=2$, $n=0.3$. As the disorder strength is increased, the magnetic moment is seen to decrease at all temperatures. Thus the enhancement of the ferromagnetic susceptibility at low temperatures cannot be explained by an increase in the number of magnetic moments with disorder. We also emphasize that the low temperature increase in $\chi_{F}$ is different from the Stoner criterion for ferromagnetism, $UN(E_F) > 1$, where $N(E_F)$ is the electron density of states at the Fermi level. Since increasing disorder (at fixed $U$) inevitably leads to loss of spectral density at the Fermi level, the Stoner criterion predicts that ferromagnetic tendencies are acted against by increasing disorder. 

\begin{figure}[hb]
\includegraphics[clip=true,angle=-90,width=0.5\textwidth]{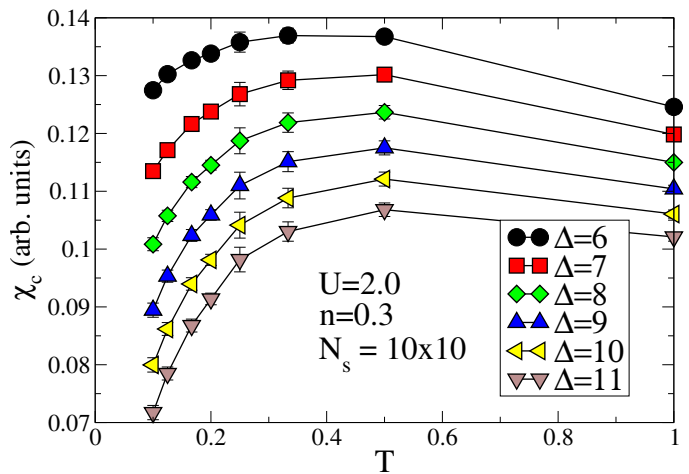}
\caption{(Color online) The charge susceptibility $\chi_{c}$ {\emph{vs}}. $T$ for different disorder strengths for $U=2$, $n=0.3$. The number of disorder realizations is the same as Fig.~\ref{U0_AF_uniform}. Note that $\chi_{c}$ is related to the compressibility $\kappa$ through $\kappa$ = $\chi_{c}/\langle n\rangle ^{2}$. Since $\langle n\rangle$ is held constant for all temperatures and disorder strengths, $\kappa$ and $\chi_{c}$ are proportional. }
\label{kappa}
\end{figure}

In Fig.~\ref{kappa}, we plot the charge susceptibility  $\chi_{c}= \partial \langle n\rangle/\partial \mu$ {\emph{vs}}. $T$ for different disorder strengths for $U=2$, $n=0.3$. The charge susceptibility decreases monotonically with increasing disorder strength for all temperatures investigated. Since the charge susceptibility  is expected to be proportional to the density of states at the Fermi level (at fixed interaction strength and electron density), we infer that $N(E_F)$ is decreasing with disorder. The charge susceptibility does not show any discernible change in behaviour with disorder at high or low temperatures, as in the case of the dc conductivity $\sigma_{\rm{dc}}$ \cite{Chakraborty:2010} but in contrast with $\chi_{F}$ and $\langle m_z^2\rangle$, thus implying that at low temperatures the charge and spin sectors display different behaviour in the simultaneous presence of disorder and interaction.        

\section{Discussion}

In conclusion, our results show that in a strongly correlated system at low enough temperatures, an increase in randomness 
can increase the tendency towards ferromagnetism. This tendency towards ferromagnetism is present for all interaction strengths we have investigated, although, as expected, stronger interaction increases the ferromagnetic susceptibility. We have shown that this enhancement in ferromagnetic suscpetibility is not due to (i) the formation of local moments and (ii) the Stoner instability. It is a ubiquitous low-temperature phenomenon in the simultaneous presence of disorder and interaction. In particular, this is not related to the phase transition between the metallic and Anderson-insulating phase. Rather, the enhancement is seen in both the metallic and insulating phases. We have also established that at low temperatures, the response of the spin sector to an increase in the disorder strength is markedly different to that of the charge sector. Similar effects are also seen in calculations using DMFT.\cite{Chakraborty:2011} Therefore Anderson localization cannot explain these effects since it is absent in $d=\infty$ where DMFT is exact. The details of the DMFT calculations will be addressed in a future publication.

This work was supported in part by the Deutsche Forschungsgemeinschaft through the TRR 80. One of us (KB) acknowledges support by the Foundation for Polish Science
(FNP) through TEAM/2010-6/2 project co-financed by the EU European Regional Development Fund. We are grateful to A. P. Kampf for helpful discussions and to M. Sekania for help with the computer cluster at the University of Augsburg.



\begin{thebibliography}{35}


\bibitem{Lee85}
P. A. Lee and T. V. Ramakrishnan, {\rm{Rev. Mod. Phys.}}, {\bf{57}}, 287 (1985).


\bibitem{Altshuler85} B.L.~Altshuler and A.G.~Aronov, in {\em
    Electron-Electron Interactions in Disordered Systems},
  eds. M.~Pollak and A.L.~Efros (North-Holland, Amsterdam, 1985),
  p.1.


\bibitem{Belitz94}  D.~Belitz and T.~R.~Kirkpatrick,
  Rev. Mod. Phys. {\bf 66}, 261 (1994).

\bibitem{Anderson50-2010} \emph{50 Years of Anderson Localization}, ed. E. Abrahams (World Scientific, Singapore, 2010).

\bibitem{Bloch:2008}
 I.~Bloch, J.~Dalibard, and W.~Zwerger,
  Rev. Mod. Phys. {\bf 80}, 885 (2008).

\bibitem{Aspect:2009}
 A. Aspect and M. Inguscio, Physics Today \textbf{62}, No. 8 (August), 30 (2009).

\bibitem{Sanchez-Palencia:2010}
 L. Sanchez-Palencia and M. Lewenstein, Nature Phys. \textbf{6}, 87 (2010).


\bibitem{Mott90} 
N.~F.~Mott, {\em Metal--Insulator Transitions}, 2nd
  edn. (Taylor and Francis, London 1990).

\bibitem{Anderson58} P.~W.~Anderson, Phys. Rev. {\bf 109}, 1492 (1958).


\bibitem{Spin_diffusion}
D.~Belitz and T.~R.~Kirkpatrick, Phys. Rev. Lett., {\bf{63}}, 1296 (1989); T.~R.~Kirkpatrick and D.~Belitz, Phys. Rev. B., {\bf{41}}, 11082 (1990); D.~Belitz and T.~R.~Kirkpatrick, Phys. Rev. B., {\bf{44}}, 955 (1991);

\bibitem{Finkelshtein83} A.~M.~Finkel'stein, Sov. Phys. JETP {\bf 75},
  97 (1983).










\bibitem{Ulmke:1997} M.~Ulmke and R.~T.~Scalettar, Phys. Rev. B {\bf 55}, 4149 (1997).

\bibitem{Ulmke:1995} M. Ulmke, V. Jani\v{s}, and D. Vollhardt, Phys. Rev. B {\bf 51}, 10411 (1995).

\bibitem{Byczuk:2003} 
K.~Byczuk, M.~Ulmke, and D.~Vollhardt, {\rm{Phys. Rev. Lett.}}, {\bf{90}}, 196403 (2003).

\bibitem{Byczuk:2009}
K.~Byczuk, W.~Hofstetter, and D.~Vollhardt, {\rm{Phys. Rev. Lett.}}, {\bf{102}}, 146403 (2009).

\bibitem{Tusch:1993}
M. A. Tusch and D. E. Logan, {\rm{Phys. Rev. B}}, {\bf{48}}, 14843 (1993).
\bibitem{Heidarian:2004}
D. Heidarian and N. Trivedi, {\rm{Phys. Rev. Lett.}}, {\bf{93}}, 126401 (2004).

\bibitem{Kravchenko:1994}
S. V. Kravchenko, G. V. Kravchenko, J. E. Furneaux, V. M. Pudalov, and M. D'Iorio, {\rm{Phys. Rev. B}}, {\bf{50}}, 8039 (1994).

\bibitem{Shashkin:2001}
A. A. Shashkin, S.V. Kravchenko, V. T. Dolgopolov, and T. M. Klapwijk, Phys. Rev. Lett. {\bf{87}}, 086801 (2001); A. A. Shashkin, S. Anissimova, M. R. Sakr, S.V. Kravchenko, V. T. Dolgopolov, and T. M. Klapwijk, Phys. Rev. Lett. {\bf{96}}, 036403 (2006).

\bibitem{Pudalov:2002}
V. M. Pudalov {\emph{et al}}., Phys. Rev. Lett, {\bf{88}}, 196404 (2002).

\bibitem{Prus:2003}
O. Prus, Y. Yaish, M. Reznikov, U. Sivan, and V. Pudalov, Phys. Rev. B., {\bf{67}}, 205407 (2003). 

\bibitem{Paalanen:1986}
M.~A.~Paalanen, S.~Sachdev, R.~N.~Bhatt, and A.~E.~Ruckenstein, Phys. Rev. Lett., {\bf{57}}, 2061, 1986.

\bibitem{Castellani:1984}
C.~Castellani, C.~Di Castro, P.~A.~Lee, M.~Ma, S.~Sorella, and E.~Tabet, Phys. Rev. B., {\bf{30}}, 1596 (1984).

\bibitem{Punnoose:2005}
 A.~Punnoose and A.~M.~Finkel'stein, {\rm{Science}}, {\bf{310}}, 289 (2005).

\bibitem{Denteneer:1999}
P. J. H. Denteneer, R. T. Scalettar, and N. Trivedi, {\rm{Phys. Rev. Lett.}}, {\bf{83}}, 4610 (1999). 

\bibitem{Denteneer:2001}
P. J. H. Denteneer, R. T. Scalettar, and N. Trivedi, {\rm{Phys. Rev. Lett.}}, {\bf{87}}, 146401 (2001).

\bibitem{Denteneer:2003}
P. J. H. Denteneer and R. T. Scalettar, {\rm{Phys. Rev. Lett.}}, {{\bf{90}}}, 246401 (2003).

\bibitem{Chakraborty:2007}
P. B. Chakraborty, P. J. H. Denteneer, and R. T. Scalettar, {\rm{Phys. Rev. B}}, {\bf{75}}, 125117 (2007).

\bibitem{Chakraborty:2010}
P.~B.~Chakraborty,  K.~Byczuk, and D.~Vollhardt, {\rm{Phys. Rev. B}}, {\bf{84}}, 035121 (2011).

 
\bibitem{Blankenbecler:1981}
R. Blankenbecler, D. J. Scalapino, and R. L. Sugar, {\rm{Phys. Rev. D}}, {\bf{24}}, 2278 (1981).



\bibitem{Varney:2009}
C. N. Varney, C. R. Lee, Z. J. Bai, S. Chiesa, M. Jarrell, and R. T. Scalettar, {\rm{Phys. Rev. B}}, {\bf{80}}, 075116 (2009).

\bibitem{Chakraborty:2011}
P. B. Chakraborty, K. Byczuk, and D. Vollhardt (unpublished).



\end{thebibliography}
\end{document}